\documentclass[11pt]{article}
\usepackage{erice,epsfig}

\def\Journal#1#2#3#4{{#1} {\bf #2}, #3 (#4)}

\def\NCA{\em Nuovo Cimento}

\def\NIMA{{\em Nucl. Instrum. Methods} A}
\def\NPB{{\em Nucl. Phys.} B}
\def\PLB{{\em Phys. Lett.}  B}

\def\PRA{{\em Phys. Rev.} A}
\def\PRC{{\em Phys. Rev.} C}
\def\PRD{{\em Phys. Rev.} D}
\def\PR{{\em Phys. Rev.}}

\def\ZP{{\em Z. Phys.}}
\def\PREP{{\em Physics Reports}}
\def\PPNP{{\em Prog. Part. Nucl. Phys.}}
\def\JPG{{\em J. Phys.} G}
\def\EPJC{{\em E. Phys. J.} C}

\def\be{\begin{equation}}
\def\ee{\end{equation}}
\def\bea{\begin{eqnarray}}
\def\eea{\end{eqnarray}}

\begin{document}
\vspace*{4cm}
\title{COHERENT ELECTROMAGNETIC PROCESSES IN RELATIVISTIC HEAVY ION COLLISIONS}

\author{ K. HENCKEN }

\address{Departement f\"ur Physik und Astronomie, Institut f\"ur Physik,\\
Universit\"at Basel, Klingelbergstr. 82, 4056 Basel, Switzerland}

\maketitle\abstracts{
Using the strong electromagnetic fields in peripheral heavy ion collisions
gives rise to a number of interesting possibilities of applications in 
both photon-photon and photon-hadron physics.
We look at the theoretical foundations of the equivalent photon 
approximation and the specific problems in the heavy ion case.
The interesting physics processes that can be studied
in this way are outlined. Electron positron pair production
plays a special role. We look at multiple pair production and Coulomb 
corrections as typical strong field effects. But electron positron 
pair production is also an important loss process and has some practical 
applications.
}

\section{Introduction}
Central collisions are the reason, why heavy ion colliders have been or
are currently built. But as soon as these colliders are available it is
of interest to look also at very peripheral (``ultraperipheral'') collisions, 
which are 
characterized by the condition that the impact parameter $b$ is larger than
the sum of the nuclear radii. The two ions do not interact directly via
the nuclear interaction in this case and only the electromagnetic field or
the long range part of the nuclear interaction (described by Pomeron-
or meson-exchange) is still able to interact. 

Peripheral collisions are quite complementary to the central ones, both
in terms of the characteristics of the events, but also concerning the physics
of interest. Peripheral collisions can be characterized as being 
``{\it silent}'' compared to the ``{\it violent}'' central interaction with 
their large multiplicities. Whereas in central collisions one is looking for 
``{\it matter under extreme conditions}'' the focus of peripheral collisions 
is more towards the ``{\it elementary interaction of photons}''. Peripheral 
collisions can in many cases study properties which cannot be easily extracted
from the central collisions.

The coherent action of all the protons within the nucleus gives an enhancement
factor of $Z^2$ or $Z^4$, which is the reason for the often very large cross 
sections and rates. The strong electromagnetic fields have already found a 
number of applications in atomic physics and in nuclear physics. In the past 
years their use at higher energies for both photon-photon and photon-hadron 
physics has been studied theoretically. Some reviews of the field can be 
found in~\cite{BertulaniB88,KraussGS97,BaurHT98}, the most recent one
is~\cite{BaurHTS01}. With the recent 
measurements at the STAR detector at RHIC this field has now left the purely 
theoretical stage and has come into reality.

\begin{figure}
\begin{center}
\psfig{figure=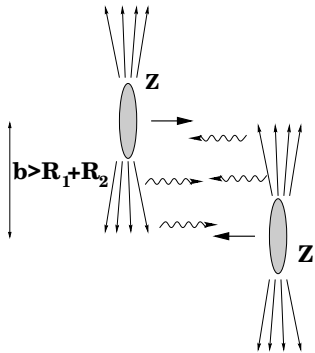,width=0.25\hsize}
\end{center}
\caption{The strong electromagnetic fields surrounding the heavy ions
in relativistic collisions are an intense source of (quasireal) photons.
They can be used in very peripheral collisions for photon-photon and 
photon-nucleus collisions.
}
\end{figure}

\section{The Equivalent Photon Approximation in the heavy ion case}
Lepton (electron) colliders have been traditionally used to study 
$\gamma\gamma$
physics. The equivalent photon approximation is one of the main 
theoretical tools there. This method was originally developed by 
Fermi~\cite{Fermi24} from the observation that the Coulomb field of a fast 
moving
object resembles the flux of a spectrum of real photons and later extended
to the relativistic case by Weizs\"acker~\cite{Weizsaecker34} and 
Williams~\cite{Williams34}. Its application for photon-photon physics is 
discussed in detail in~\cite{BudnevGM75}. The main idea behind the quantum
mechanical derivation of the equivalent photon approximation is the fact
that photons with a small virtuality $q^2$ are dominant due to the singular
behavior of the photon propagator. The cross section for
virtual photons can then be replaced by the one for real photons
and the cross section is written as
\be
\sigma(ee \rightarrow ee X) = \int \frac{d\omega_1}{\omega_1}
\frac{d\omega_2}{\omega_2} n(\omega_1) n(\omega_2) 
\sigma(\gamma\gamma\rightarrow X),
\ee
where $n(\omega)$ is the equivalent photon number in the photon-photon
case. 

\begin{figure}
\begin{center}
\psfig{figure=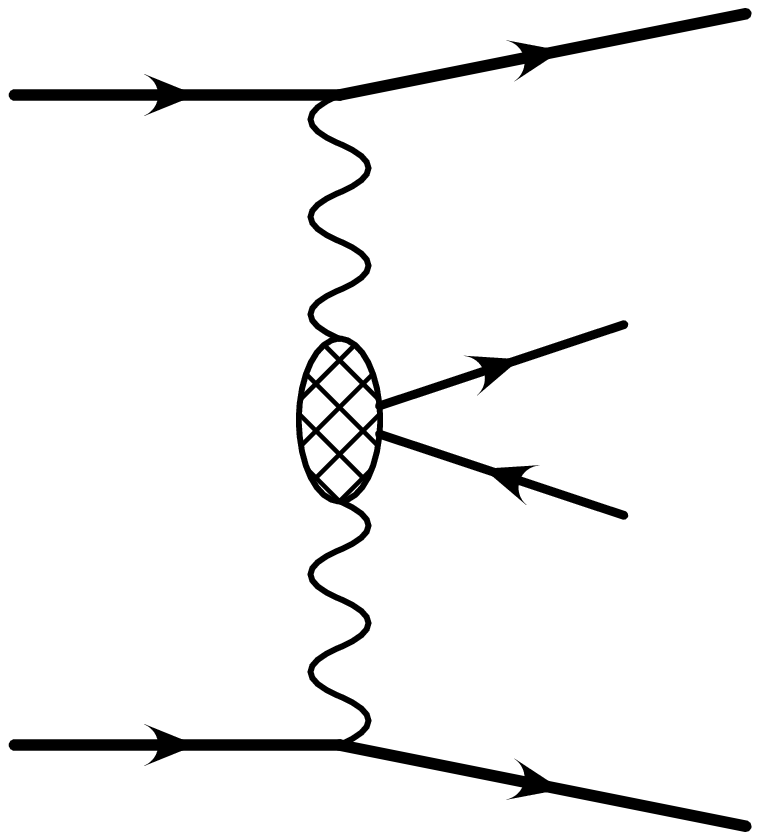,width=0.15\hsize}~(a)~~~
\psfig{figure=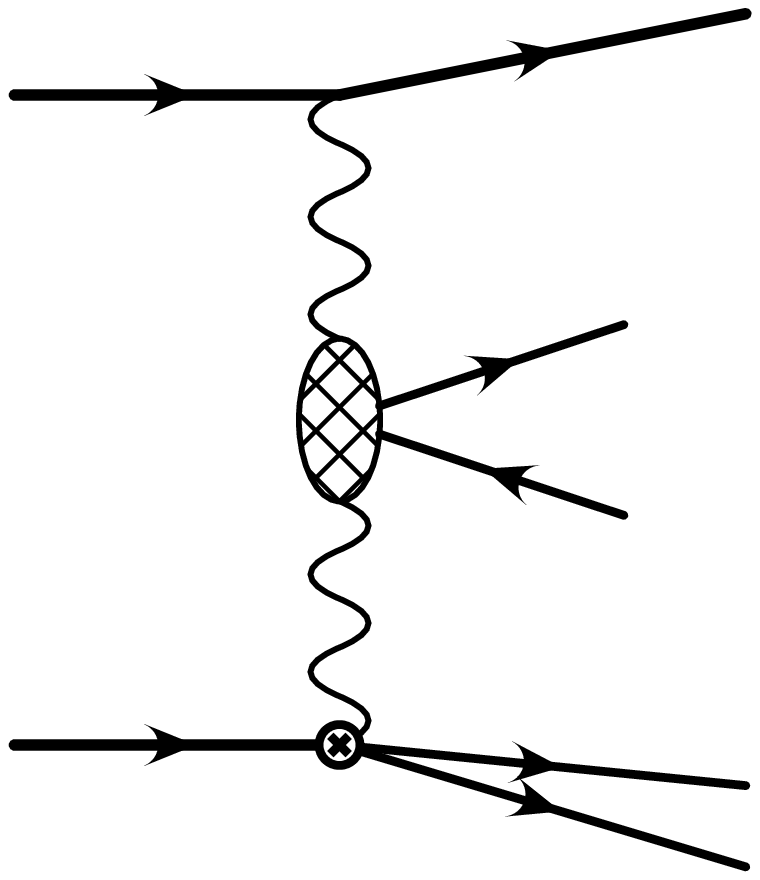,width=0.15\hsize}~(b)~~~
\psfig{figure=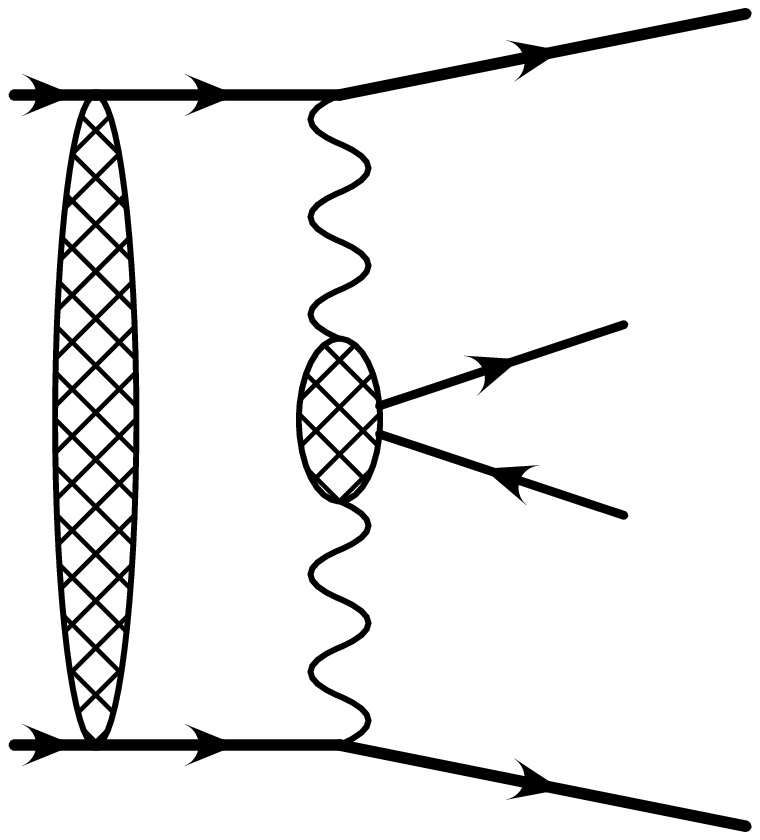,width=0.15\hsize}~(c)~~~
\psfig{figure=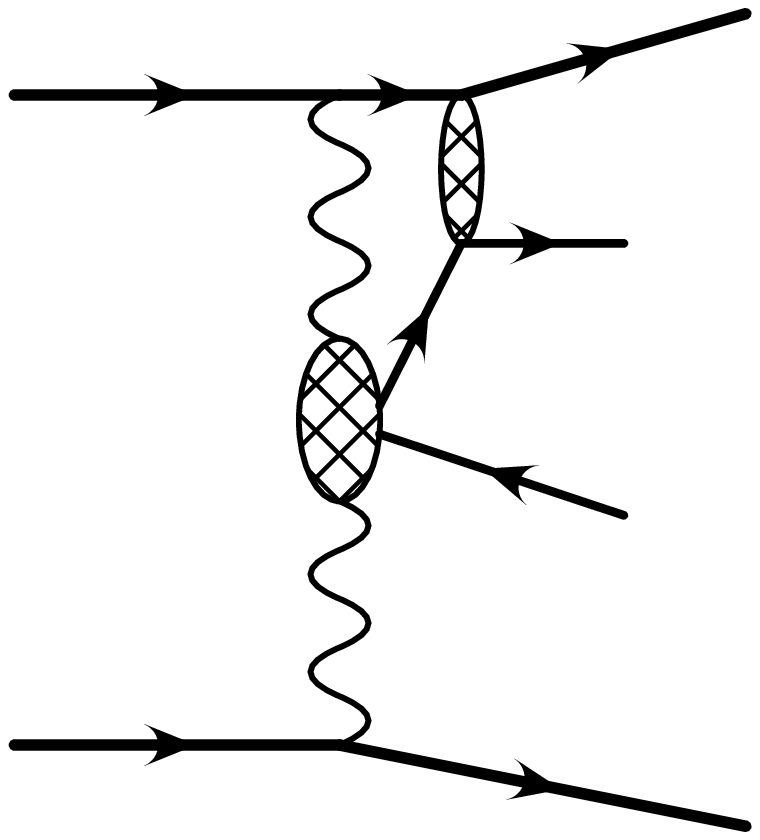,width=0.15\hsize}~(d)
\end{center}
\caption{Due to the more complex nature of the ions corrections
occur in photon-photon processes in the heavy ion case. Besides the elastic
photon emission, which is governed by the elastic form factor (a), the
photon can also be emitted inelastically (b). The two ions of course do
interact with each other either via the Coulomb interaction or
due to their nuclear interaction for small impact parameter (c) (``initial
state interaction''). Finally if a hadronic final state is produced it can
again interact with one of the ions (d) (``final state interaction'').
\label{fig:1}}
\end{figure}

More recently photon-photon physics was also investigated for 
$ep$ collisions or $eA$ collisions at HERA~\cite{herafuture96}. In ion-ion
collisions there are current measurements at STAR and both CMS and ALICE have
plans to measure photon-photon and photon-hadron processes as well. In 
contrast to the point-like structure-less electrons with a small charge 
and only electromagnetic interaction the situation with ions is more complex. 
A number of additional processes, which occur in the photon-photon case, are 
shown in Fig.~\ref{fig:1}.
Similar diagrams exist as well in the photon-ion case. Let us address the
different processes in more detail.

The finite size of the ions can be taken care of by using the elastic form
factor. The inelastic excitation will contribute also to the photon spectrum.
This was studied for the excitation of the most important excited state, the
giant dipole resonance (GDR), in~\cite{HenckenTB96}. It was found that this is 
only a rather small contribution, whereas, e.g., the $p$-$\Delta$ transition 
gives an additional effect of about 10\%. At even higher $q^2$ one also has 
incoherent contributions. For the proton it was found that the photon 
emission from individual quarks gives a contribution that is larger than the 
elastic one~\cite{OhnemusWZ94,DreesZ89}. In the heavy ion case such a 
contribution is expected to be much smaller for two reasons: First the charge 
of the ``partons'' (quarks for the proton, protons for the heavy ion) is small
compared to the total charge for ions, whereas they are of the same order for
the proton. Therefore the loss of the coherence enhancement will be larger
here. In addition the exclusion of the initial/final state interaction (see
below) will remove a large part of this contribution. The virtuality of the
photon in the inelastic case needs to be rather large, $q^2\gg 1/R^2$; their
range will therefore be rather small. On the other hand the exclusion of the
initial/final state interaction eliminates most of the small impact parameters.

The exclusion of the nuclear interaction at small impact parameter 
(see Fig.~\ref{fig:1}(c)) is not
easy to do in the plane wave approach. Of course photon-photon
processes will occur in this case as well but they are completely dominated
by the particles produced in the nuclear interactions; therefore these 
collisions are not useful and one should exclude them also from luminosity
calculations. A possible approach for the inclusion of the strong interaction 
is a calculation starting from the eikonal 
approximation, see also the discussion in~\cite{Gerhard}. The strong Coulomb 
interaction between the two ions, characterized by the large Sommerfeld 
parameter $\eta\approx Z_1 Z_2\alpha\gg1$, on the other hand allows to use the
semiclassical approximation, see also~\cite{Gerhard}. At high energies the 
heavy ions move essentially along a straight
line and it is possible to define an impact parameter dependent photon
number. In the case of a point-like particle this number is well known to be
\be
N(\omega,b) = \frac{Z^2\alpha}{\pi^2 b^2} \left(\frac{c}{v}\right)^2 x^2
\left[ K_1^2(x) + \frac{1}{\gamma^2} K_0^ 2(x) \right], 
\quad x=\frac{\omega b}{\gamma}.
\ee
In such a semiclassical pictures the exclusion of the nuclear interaction can 
be done by excluding all impact parameters with $b<R_1+R_2$. Alternatively
one can instead introduce the probability for no hadronic interaction given,
e.g., from Glauber theory and get the photon-photon flux by integrating the
product of the equivalent photon numbers and this probability over all
impact parameter. In general the difference between this two approaches is
not very large.

\begin{figure}
\begin{center}
\psfig{figure=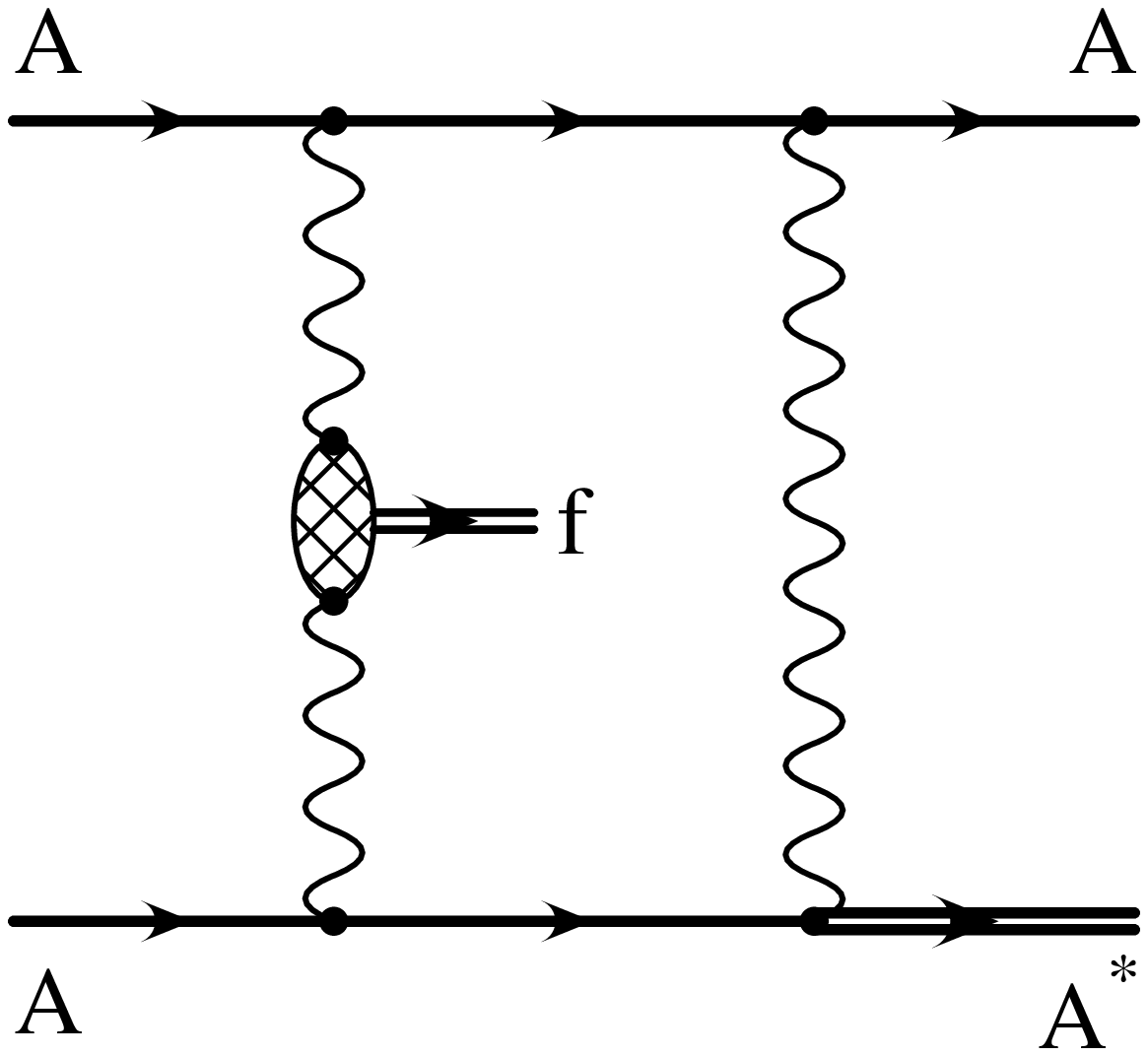,width=0.2\hsize}~~~~~~
\psfig{figure=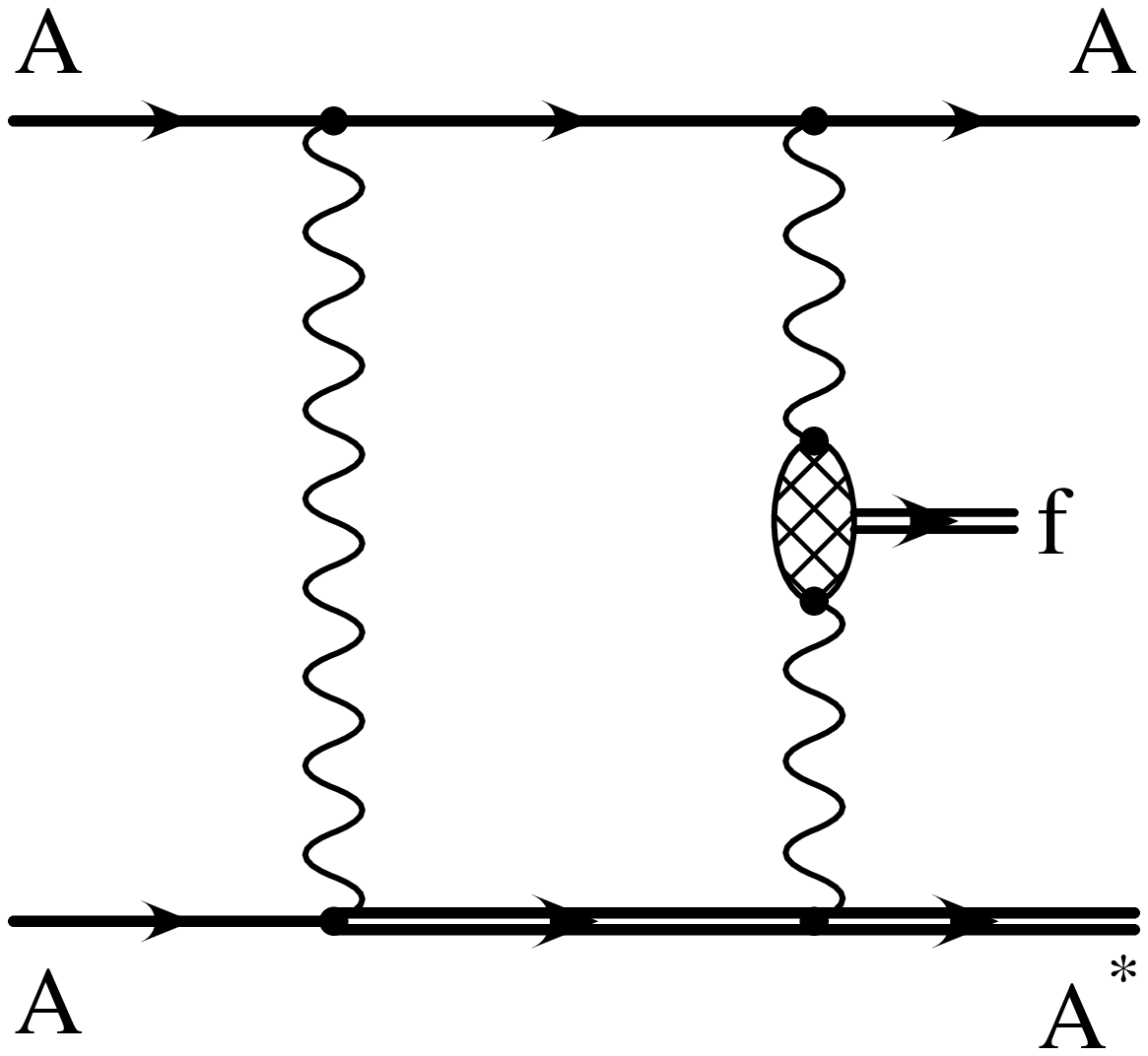,width=0.2\hsize}
\end{center}
\caption{The strong fields allow for the excitation of the ions in addition
to the photon-photon process. For the heaviest ion the probability for this
is rather large. The semiclassical approximation allows for a very
easy description of these processes in terms of the products of the individual
probabilities.
\label{fig:addproc}}
\end{figure}

Another type of ``initial state interaction'' are additional
electromagnetic processes occurring together with the  photon-photon or 
photon-hadron interaction (Fig.~\ref{fig:addproc}). In the semiclassical 
approximation this process will be described by the product of the individual 
probabilities, which is
discussed also in~\cite{Gerhard}. The main contribution is
expected to be from the GDR, but excitation to higher energies contribute as
well. For the heaviest ion the probability for GDR excitation is rather large 
and approaches one. Therefore the excitation of higher states (DGDR), 
see~\cite{Carlos}, need to be taken into account. It is found that about 95\%
of all Pb-Pb collisions are accompanied by such an excitation, but only about
1\% of the Ca-Ca collisions. As the equivalent photon number $N(\omega,b)$
falls off as $1/b^2$, a very convenient parameterization of this probability is
\be
P(A^*,b) \approx 1 - \exp\left[-(\mbox{const}/b)^2\right]
\ee
The importance of such an effect depends on
the trigger condition used in the experiment. If one triggers for no breakup
of the ions at all, one would loose this part of the luminosity. On the other
hand this effect is currently used at RHIC to trigger for peripheral collisions
by looking for the neutrons coming from the decay of the GDR in both ions, 
which are measures in both ZDCs~\cite{Klein01b}.

Due to their much larger ion luminosity, it was found
that medium heavy ions (like Ca-Ca) give in the end a much larger total
luminosity (given by the product of ion luminosity and photon-photon
luminosity). Therefore medium heavy ion beams are often more preferable.
In addition the ions have a smaller size and therefore the maximum
photon energy $\omega_{max}=\gamma/R$ will be larger in this case as well.

Apart from the proton-proton case (see also~\cite{Kristof}) it will probably
be impossible to measure the scattered ions, therefore only ``untagged 
events'' will be possible. These are 
characterized by a small total transverse momentum of the final state smaller
than about $2/R$. In terms of the rapidity the photon luminosity peaks 
at $Y=0$ and has a range of about
\be
Y_{max} \approx \ln 2\omega_{max} / M_{\gamma\gamma}.
\ee

\section{Physics Potential in $\gamma\gamma$ and $\gamma A$ collisions}
The available invariant mass range in the photon-photon case is given
by the maximum photon energy $\omega_{max}\approx \gamma/R$ and therefore
the maximum invariant mass is $2\gamma/R$~\footnote{Of course these ``maximum
energies'' should not be treated as absolute maxima. Photon with higher
energies do occur but their spectrum decreases exponentially.}.
The invariant mass range for RHIC extends up to a few GeV and 
therefore makes this of interest for meson spectroscopy in this range.
The luminosities are comparable to what has been available at
LEP. Both meson production and meson pair production can be studied.
Detailed studies of these processes have been done by the ``Ultraperipheral
Collision Group'' at STAR, see also~\cite{Joakim,Pablo}. The suppression of
the two-photon production of a meson would also be an indication for its
gluonic nature, making it possible to look, for example, for glueballs.

\begin{figure}
\begin{center}
\psfig{figure=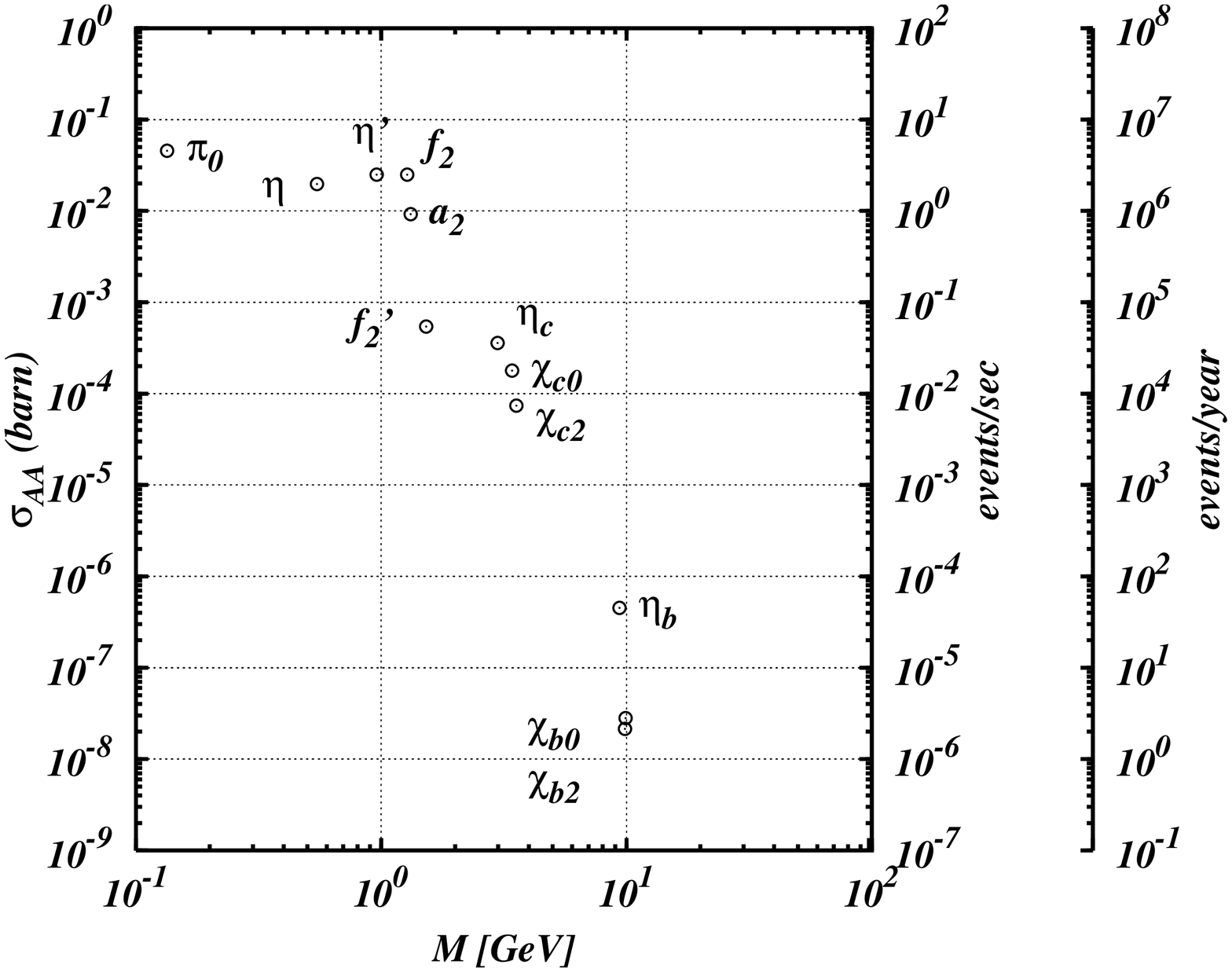,width=0.4\hsize}(a)~~~
\psfig{figure=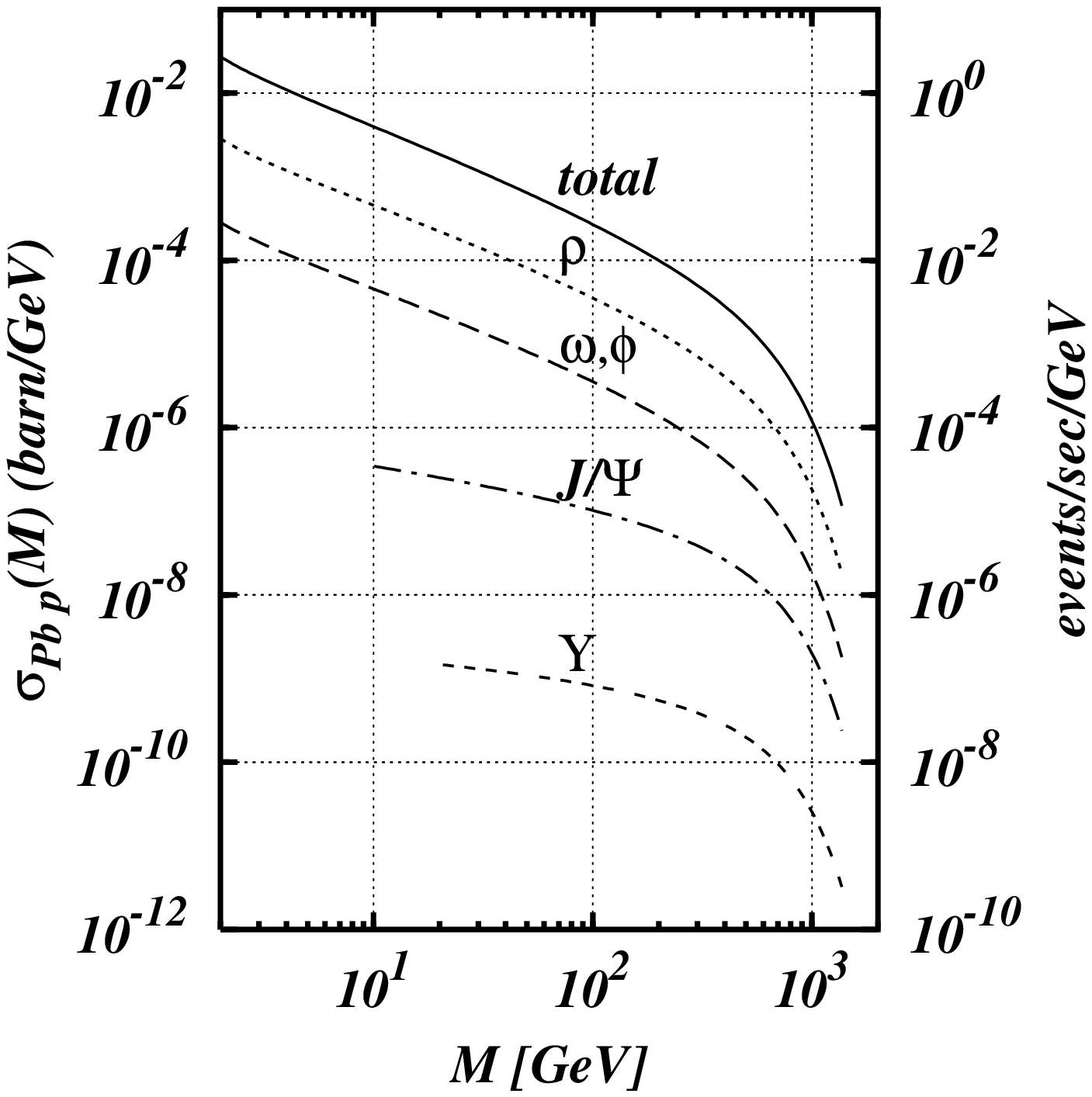,width=0.3\hsize}(b)
\end{center}
\caption{The cross sections and rates for different mesons in $\gamma\gamma$
collisions are shown for meson production in Pb-Pb collisions at the LHC in
(a). Figure (b) shows the same properties for the diffractive vector meson
production as well as the total hadronic cross section for a hypothetical
collision of Pb on protons for LHC energies. The results for Pb-Pb
collisions can be obtained by a multiplication of a factor describing the
coherent (or incoherent) production. For the rates a ion-ion luminosity of 
$10^{26}$cm$^{-2}$s$^{-1}$ was assumed.
\label{fig:ga}}
\end{figure}

For the LHC the invariant mass range goes up to a few 100 GeV and therefore
extends the possibilities of LEP both in mass range and in luminosity.
Meson spectroscopy is possible there, especially for mesons with
$b$ and $c$ quarks, see Fig.~\ref{fig:ga}(a). In addition the total cross 
section 
$\gamma\gamma\rightarrow$hadrons has been measured at LEP. Deviations from
the Pomeron universality have been found there. Photon-Photon processes
at the LHC have the potential to look for this process
at higher energies. In addition it is worthwile to study whether the detectors
there will be able to detect also in the more forward direction, where most 
of the events will occur and which up to now are undetectable and need to be 
modeled.

The possibility to find new particles at the LHC has attracted some interest
in the past. Especially Higgs boson production and the search for 
supersymmetric particles have been studied in detail. Unfortunately the 
production rates (for ion-ion collisions) seem to be rather small (few per 
year) in these cases and for realistic parameters.

Photon-ion collisions are possible over a large range of energies. Please 
remember that the maximum photon energy in this case is given by 
$\omega_{max}=\gamma_{ion}/R$ with the Lorentz factor 
$\gamma_{ion}=2\gamma_{coll}^2-1$ in the rest frame of one of the ions.
Therefore the maximum photon energy is about 300 GeV for RHIC and about
500 TeV for the LHC. The total photon-ion cross section is dominated by
the nuclear excitation mainly of the GDR (the so called 
``Weizs\"acker-Williams process'')
at the lowest energies. This is an important loss process as the excited
nuclei decay mostly via neutron emission. But it is also used at RHIC
as a luminosity monitor, see~\cite{Sebastian}. Going to higher 
energies coherent diffractive processes are of interest. At the LHC
the coherent diffractive vector meson production on a nucleus up to the 
Upsilon can be studied. The expected rates for the hypothetical collisions of 
photons from Pb on a proton are shown in Fig.~\ref{fig:ga}(b). For the heavy 
ion case they need of course to be multiplied by a factor depending on the 
details of the coherent production on the nuclei. In this way one might be
able to see the transition from the ``soft'' to the ``hard'' Pomeron with
increasing meson mass. Looking for vector mesons with a larger transverse 
momenta one will also be able to study the inelastic vector meson production.
Again Fig.~\ref{fig:ga}(b) will be helpful to estimate the rates.
Finally it has been proposed to study the photon-gluon fusion process
and with it to be able to measure the in-medium gluon distribution function.
Such a measurement would be very interesting as some models predict a 
saturation of the gluon density at small Bjorken $x$.

\section{Lepton Pair Production and QED of Strong Fields}
An interesting subject in itself is the production of light lepton
pairs. Electron-positron pair production (and to some extend also muon pair
production) plays a special role in peripheral heavy ion collisions. Due to 
their small mass they are produced quite easily. It is interesting to note 
that the work of Landau and Lifschitz in 1934~\cite{LandauL34}, followed by
Racah 1937~\cite{Racah37}, are probably the first calculation of an 
electromagnetic process in relativistic ion collisions. The cross section
for $e^+e^-$ pair production is quite large. Using, e.g., the formula
derived by Racah~\cite{Racah37}
\begin{equation}
\sigma = \frac{Z^4\alpha^4}{\pi m^2} \frac{28}{27} \left[ \ln^3\gamma_{ion}^2
- 2.19 \ln^2\gamma_{ion}^2 + \cdots \right],
\end{equation}
one finds cross sections of about 30kbarn for Au-Au collisions at RHIC and 
200kbarn for Pb-Pb collisions at LHC, leading, e.g., to $10^7$ pairs/sec
produced in Pb-Pb collisions at LHC. This cross section increases with 
$\ln^3\gamma_{ion}^2$, which is rather fast and will eventually lead to a violation
of the Froissart bound. Therefore new phenomena will occur at high energies.

Most electrons and positrons are produced with rather modest energies (of
the order of several $m_e$) and into the very forward direction. Therefore
they will remain unobserved. Still the cross section for large angles and 
large energies is quite sizable. Finally we want to point out that the 
equivalent photon approximation can only be used with great care for $e^+e^-$ 
pair production. This is due to the smallness of $m_e$ compared to $1/R$.
The total cross section can be calculated within EPA by using as
cutoff parameter $m_e$ instead of $1/R$ or restricting the impact parameters
to be larger than $\lambda_c=1/m_e\approx400$fm. Calculations
using only a restricted part of the phase space or for small impact parameter
on the other hand need to be done either with a more refined analysis of this
cutoff parameter or using a full calculation.

\subsection{Multiple Pair Production and Coulomb Corrections}
\begin{figure}[tb]
\begin{center}
\begin{tabular}{cc}
\psfig{figure=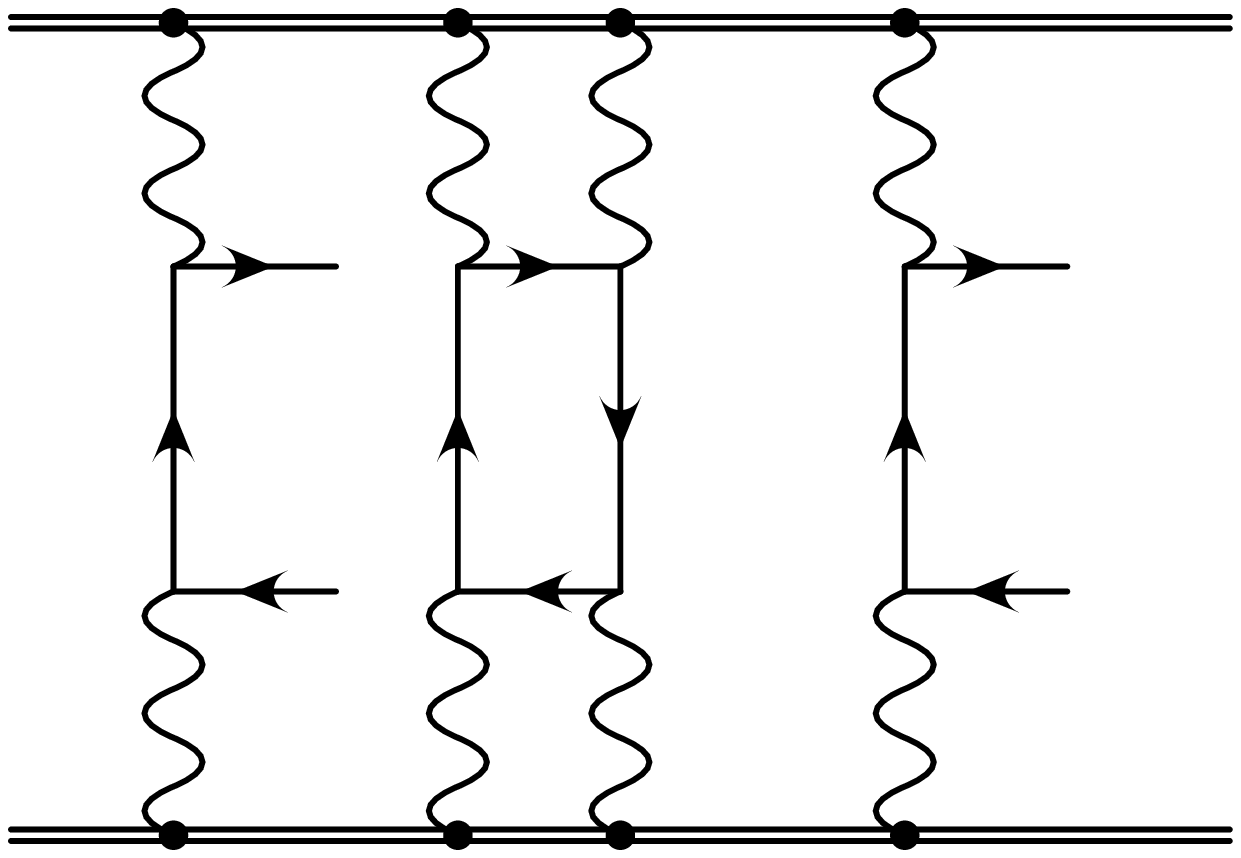,width=0.14\hsize}&
(a)\\
\psfig{figure=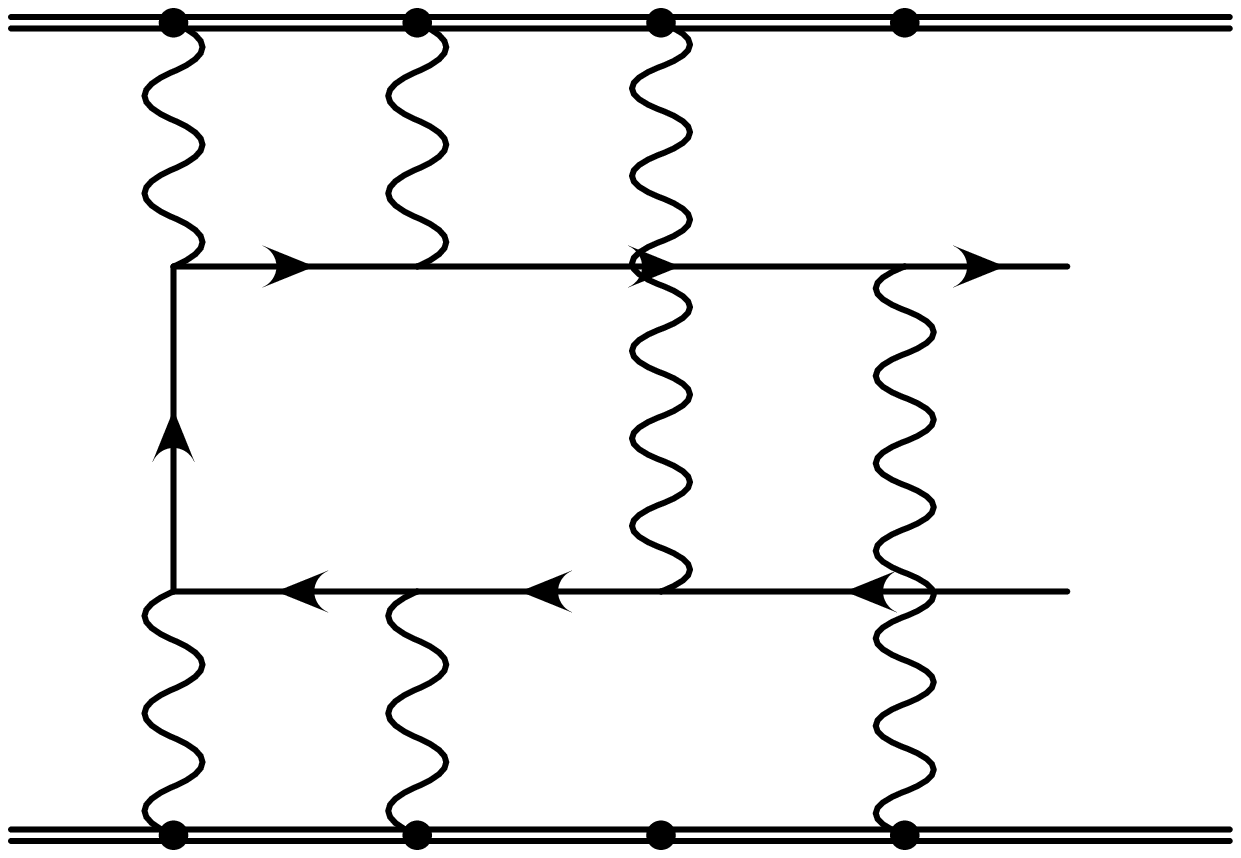,width=0.14\hsize}
&(b)\\
\psfig{figure=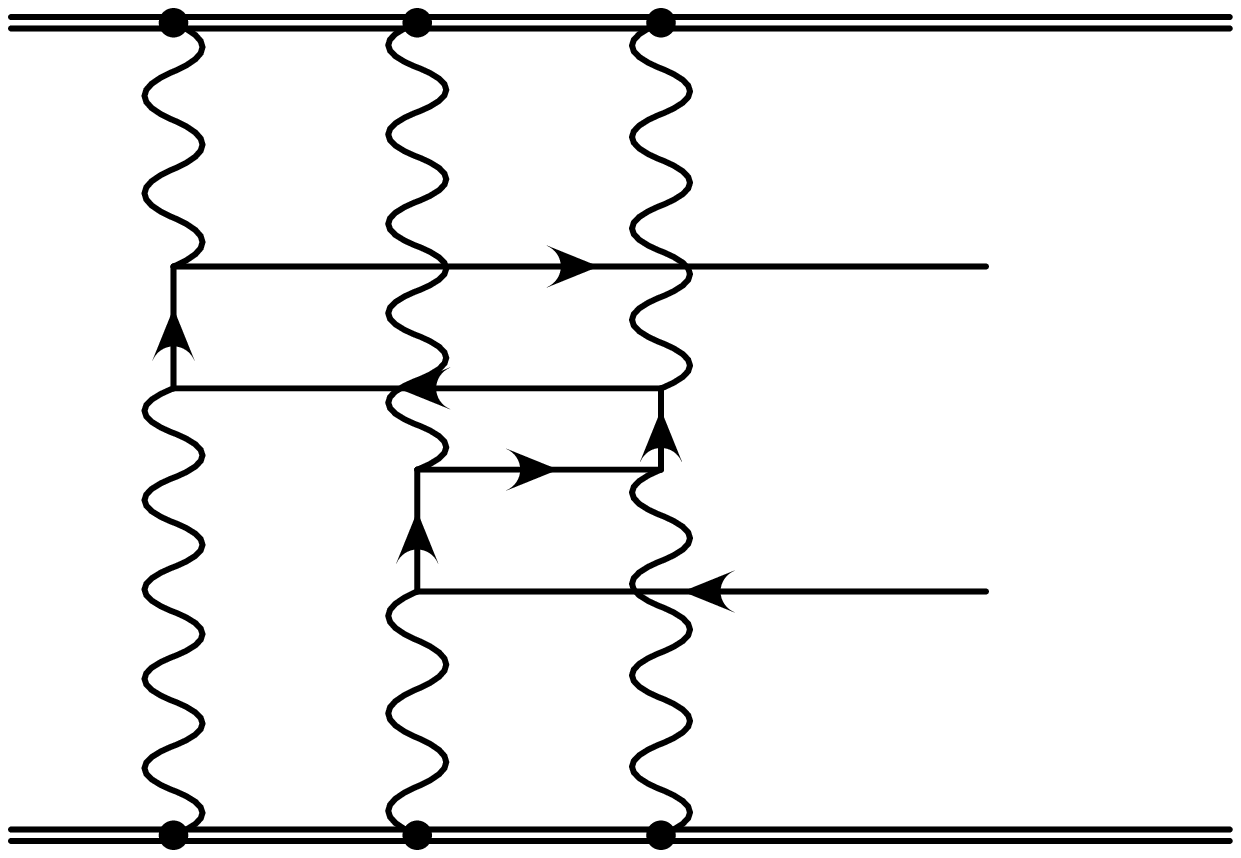,width=0.14\hsize}
&(c)
\end{tabular}
\begin{tabular}{cc}
\psfig{figure=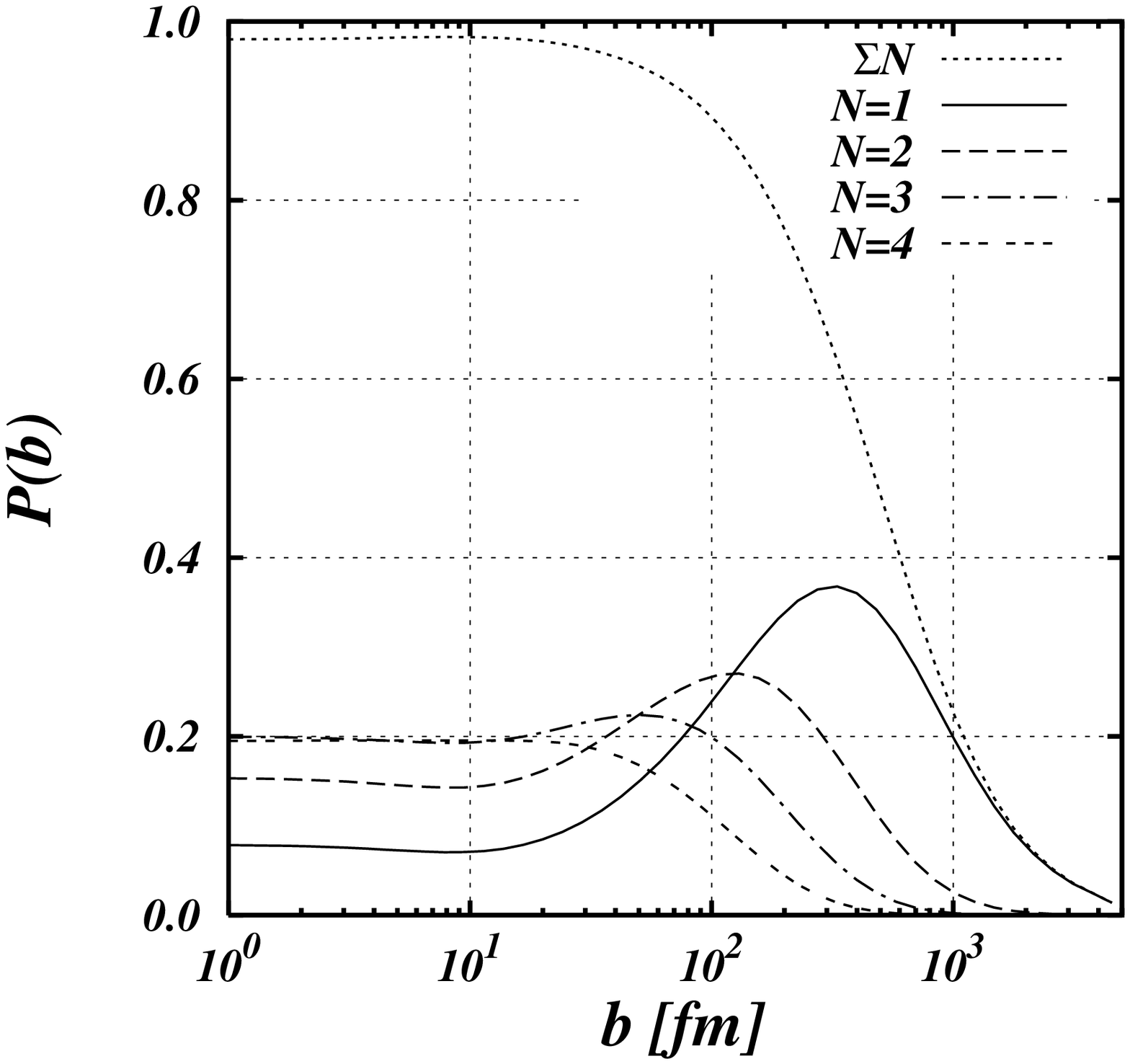,width=0.33\hsize}
&(d)
\end{tabular}
\end{center}
\caption{The electromagnetic fields in relativistic heavy ion collisions
are so strong that higher order processes will occur. Multiple pair production
in a single collision is possible as shown in (a). Coulomb corrections lead 
to the distortion of the wave functions of electron and positrons (b). 
Finally due to the strong fields also multiple-particle processes can occur 
(c). The impact parameter dependent probability for the production of up to
four pairs in lowest order is shown in (d) for Pb-Pb collisions at the LHC.
\label{fig:multiee}}
\end{figure}

Electron-positron pair production has attracted some interest in recent years
due to the observation that the impact parameter dependent probability 
calculated in perturbation theory can exceed unity already for RHIC and for 
impact parameters up to the Compton wavelength of the electron 
($\lambda_C\approx400$fm)~\cite{Baur90}. That the pair production cross section
rises too fast with collision energy was already observed by 
Heitler~\cite{Heitler34}, but was 
thought to be an ``academic problem'' by him. In a series of papers starting 
from~\cite{Baur90} it was found that the probability larger than one means 
that more than one pair will be produced on the average within a single 
collisions (Fig.~\ref{fig:multiee}(a)). 

Neglecting the antisymmetrization of the final state (and therefore treating 
the pair as a ``quasiboson'') the probability $P(N,b)$ for the $N$ pair 
production was found to follow a Poisson 
distribution~\cite{HenckenTB95a,AlscherHT97}
\be
P(N,b) = \frac{P^N(b)}{N!} \exp(-P(b))
\ee
with $P(b)$ the probability for a single lepton line. $P(b)$, which is the
probability calculated, e.g., in lowest order perturbation theory, is naturally
interpreted as the average multiplicity of pairs
\be
\left< N\right>(b) = P(b).
\ee
Corrections to the Poisson distribution, as well as multiple-particle effects,
have been discussed in~\cite{HenckenTB95a}, where also the earlier work 
is discussed.

The impact parameter dependent probability was calculated in lowest 
order~\cite{HenckenTB95b} and it was found that on the average three to four
pairs will be produced for small impact parameter. Cross section for the 
production of up to five pairs have been given as well~\cite{AlscherHT97}
(see Fig.~\ref{fig:multiee}(d)). The results for impact parameter of the order
of a few $\lambda_c$ have been recently confirmed~\cite{LeeMS01}.

As the fields are rather strong (the effective coupling constant $Z\alpha=0.6$
is not small) one might doubt the results of the lowest order calculations, 
expecting that higher order processes (on the single pair production process)
should be important. Two different types
of higher order corrections can be loosely distinguished~\cite{Aste01}: The 
distortion of the electron and positron wave-functions after their production 
are normally subsumed under the term ``Coulomb corrections'', see also 
Fig.~\ref{fig:multiee}(b). A 
second class deals with the fact that more than one pair can be present at an 
intermediate stage before the pairs annihilate crosswise 
(Fig.~\ref{fig:multiee}(c)). Such 
``multiple-particle'' processes~\cite{HenckenTB95a} are unique to the situation
of heavy ion collisions with two strong fields. A calculation for small
impact parameter~\cite{HenckenTB95a} found them to be rather small.

A classical result for Coulomb corrections in the pair production cross 
section are the Bethe-Maximon corrections~\cite{BetheM54,DaviesBM54},
describing the production of an electron-positron pair through a single (real) 
photon interacting with a strong Coulomb field~\footnote{This coincides with
the limit $Z_1\rightarrow0$ in the heavy ion case.}. In the
Bethe-Heitler formula
\begin{equation}
\sigma = \frac{28}{9} \frac{(Z\alpha)^2}{m_e^2} \left[ \ln \frac{2 \omega}{m_e} -
\frac{109}{42} - f(Z\alpha) \right] ,
\end{equation}
the Coulomb corrections are contained in the correction term 
\begin{equation}
f(Z\alpha) = (Z\alpha)^2 \sum_{n=1}^{\infty} \frac{1}{n(n^2+(Z\alpha)^2)}  
= \gamma + \mbox{Re} \psi(1+i Z\alpha)
\end{equation}
with the Euler constant $\gamma\approx0.57721$ and $\psi$ the Psi (or Digamma)
function.

Using an analysis of the different orders in $\ln\gamma$ of the contribution
with either {\it one} or {\it multiple} photon emissions from each ion, the
Coulomb corrections to the total {\em cross section} are calculated 
in~\cite{IvanovSS99}. Rather large reductions compared to the Born cross 
sections (about 25\% and 14\% for RHIC and LHC) are given. Such an analysis 
is unfortunately not helpful for the calculation of multiple pair production 
as this requires the exchange of {\it multiple} photons from each ion.

A different approach was proposed by a number of 
authors~\cite{SegevW98,BaltzM98,EichmannRSG99} making use of the form of the 
interaction in the high energy limit $\gamma\rightarrow \infty$. Using 
retarded boundary conditions it is found that only a certain class of 
diagrams is dominant in the high energy limit. The sum to all orders can be 
related to the one in lowest order approximation by a replacement of the 
photon propagator
\be
\frac{1}{q^2} \rightarrow \frac{1}{\left(q^2\right)^{1-i Z\alpha}}
\ee
A calculation of the impact parameter dependent probability was done 
in~\cite{HenckenTB99}, where it was found that the Coulomb corrections lead
to a substantial reduction of the probabilities at small impact parameter and
therefore also of the multiple-pair production cross sections.

It was already pointed out in~\cite{BaltzM98,SegevW98b} that this high energy 
limit leads to the interesting result that the cross section (but not
$P(b)$ itself) including all higher orders is identical to that in lowest 
order. This is of course in disagreement to the Bethe-Maximon corrections as 
discussed above. Different solutions for this have been
proposed in the mean time (for a discussion see~\cite{BaurHTS01}). 
That a too simplified use of the eikonal 
approximation will lead to a result in disagreement with Bethe-Maximon
theory was already pointed out in~\cite{BlankenbeclerD87}.
Using a more refined analysis~\cite{LeeM00} the 
Bethe-Maximon corrections were found to be present here as well. Still the
Coulomb corrections for the multiple pair production process have not been
calculated up to now.

\subsection{Bound-Free Pair Production}
The electron from the produced pair has a tiny probability to be produced
not as a free particle but into the bound state of one of the ions
\be
A + A \rightarrow A + (A e^-)_{K,L,\cdots} + e^+
\ee
Even though this probability is small in combination with the large cross 
section for the pair production makes for a cross section of the order
of 100~barn. As the $Z/A$ ration changes to $(Z-1)/A$ in this process,
the ion is lost from the beam. Together with the electromagnetic excitation
of the ions themself this is the dominant loss process. But whereas the
excited ions, which subsequently decay mostly by neutron emissions 
have a large momentum spread, the momentum transfer coming 
from the electron capture is small and a very narrow Pb$^{81+}$ beam emerges 
after the interaction region. Recently it was shown~\cite{Klein01} that this 
narrow beam of ions will hit the wall hundreds of meters after the 
interaction region in a very narrow spot depositing an energy of the order
of tens of watts in a very small spot and the maximum beam luminosity of the 
Pb beam at the LHC is limited due to the magnet cooling in this region.

Recently a full calculation of the bound-free pair production in first order
in the target interaction was made not only for production into the 
$1s$-state but also to higher $s$ and $p$ states~\cite{MeierHHT01}. The 
results were also compared with other calculations up to now. In general a 
good agreement was found, therefore one expects that the size of this process 
seems to be well under control.

\subsection{Equivalent Leptons: Heavy Ions as a source of high energetic
leptons}
\begin{figure}
\begin{center}
\begin{tabular}{cc}
\psfig{figure=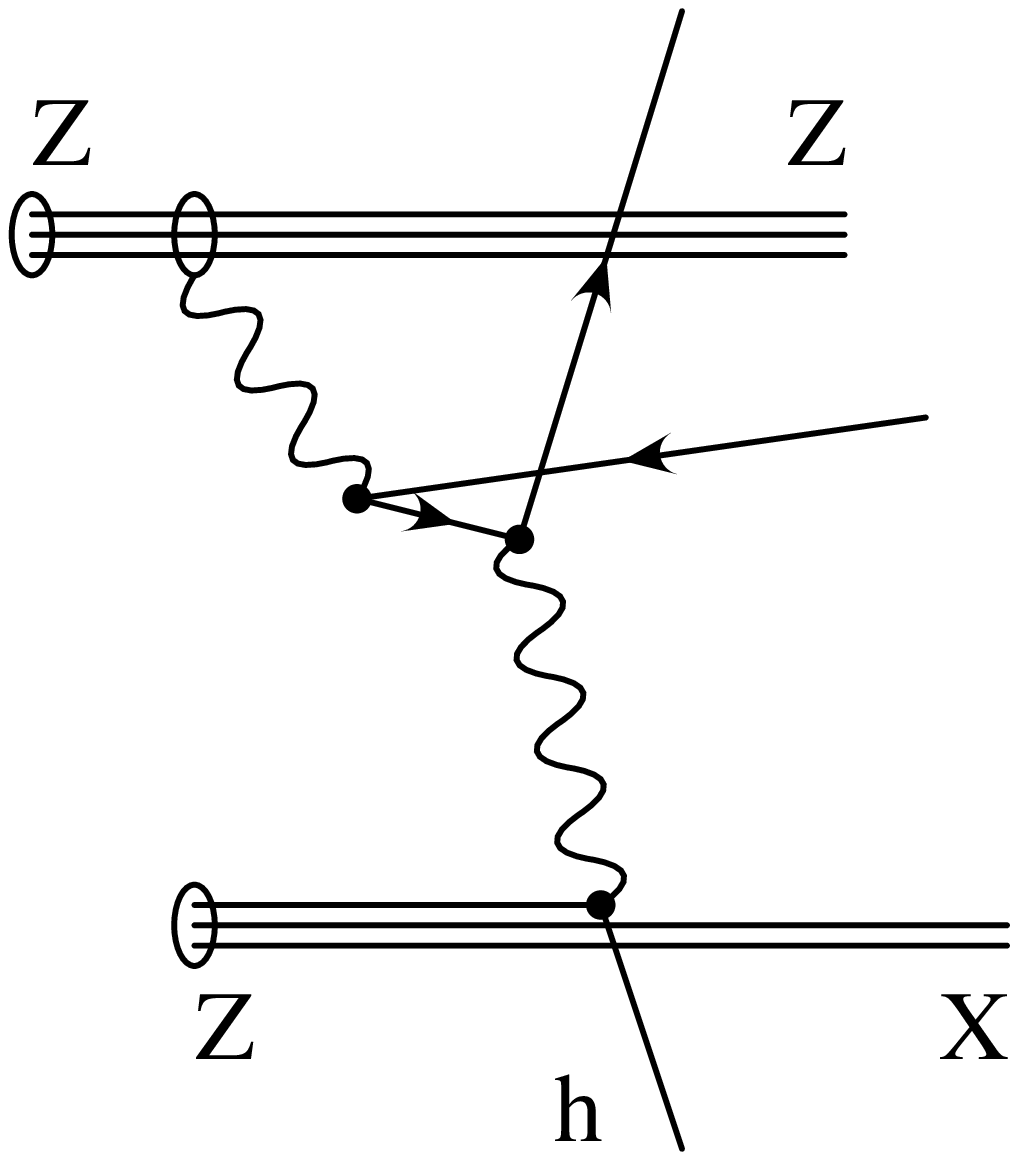,width=0.15\hsize}&
(a)\\
\psfig{figure=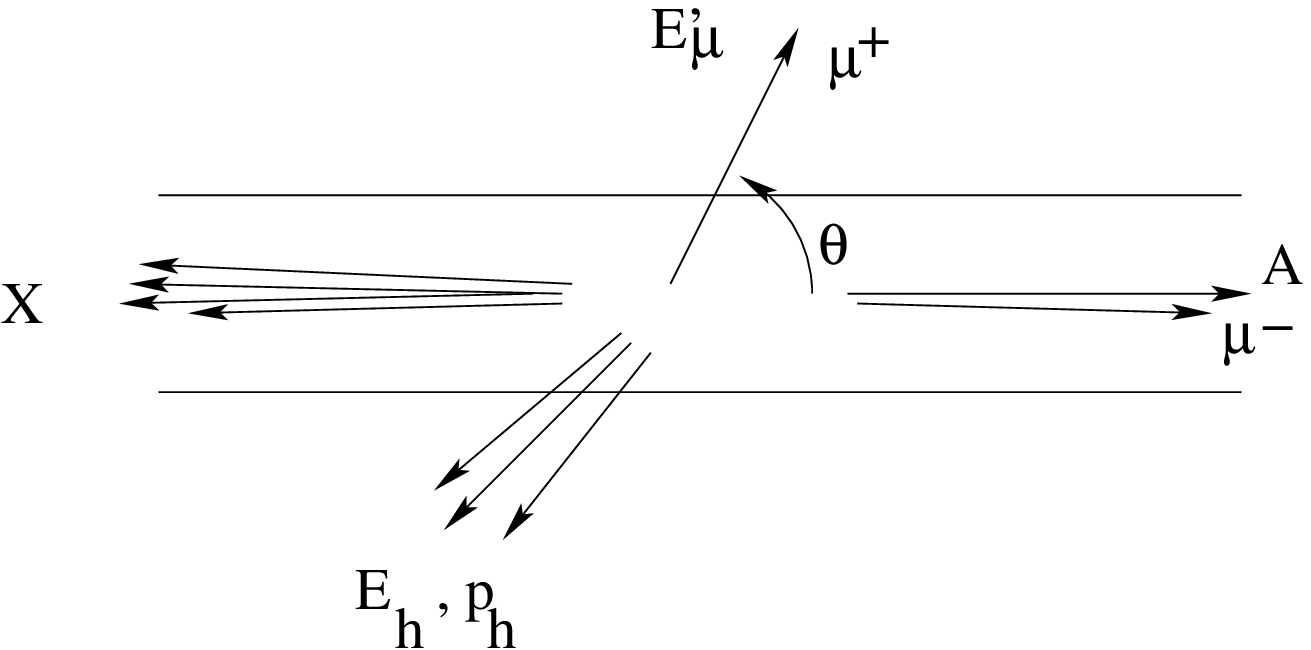,width=0.33\hsize}&
(b)
\end{tabular}
\begin{tabular}{cc}
\psfig{figure=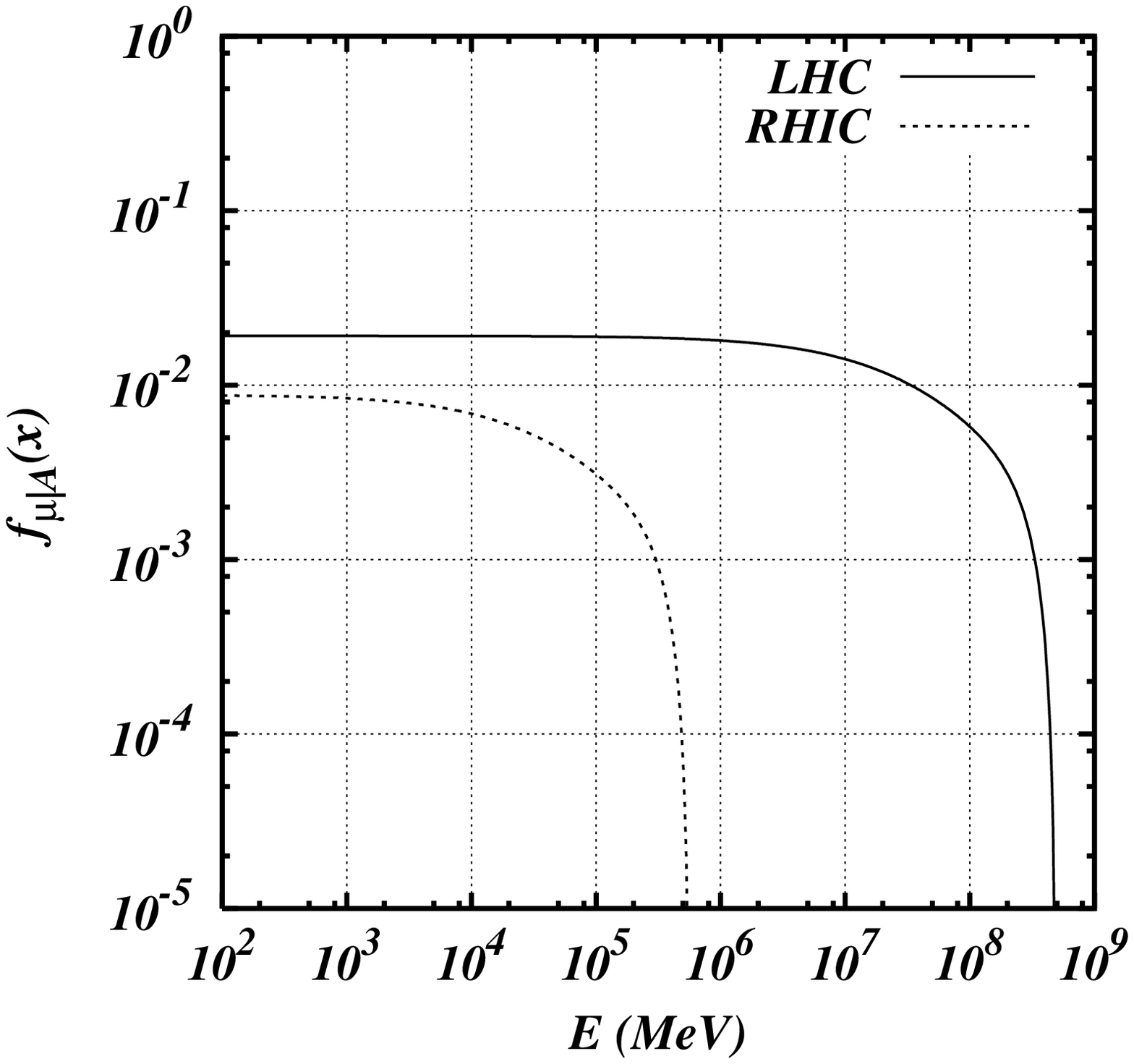,width=0.33\hsize}&
(c)
\end{tabular}
\end{center}
\caption{Processes with one muon emitted under a large angle and a jet
in the opposite direction can be thought of to occur from the deep
inelastically scattering of an equivalent muon (a). They are characterized by
one muon scattered into one direction and a jet in the opposite one (b).
The spectrum of the muons as a function of their energy is shown for both 
Pb-Pb collisions at LHC and Au-Au collisions at RHIC (c).
\label{fig:ema}}
\end{figure}

The process of lepton (in the following mainly muon) production with
a large transverse momentum of one of the muons can be best described
within the picture of the ``equivalent lepton 
approximation''~\cite{ChenZ75,BaierFK73}. The photon emitted from one of
the ions contains muons as partons, which can scatter deep inelastically
with the other ion (see Fig.~\ref{fig:ema}(a)). Using this approach the spectrum of the
equivalent muons (or electrons) can be written as
\be
f_{\mu|Z}(x) = \int_x^{1/(m_A R)} du f_{\gamma|Z}(u) f_{\mu|\gamma}(x/u).
\ee
With this the cross section can be calculated, see Eq.~(144)
of~\cite{BaurHTS01}. The equivalent muon spectrum for RHIC and LHC are 
given in Fig.~\ref{fig:ema}(c).
\begin{figure}
\end{figure}

The deep inelastic scattering process will give information about
the quark distribution function within the nucleus. In contrast to a
lepton beam, which is mono-energetic, the muon here has a continuous spectrum.
Nevertheless by measuring the scattered muon and the energy and 
longitudinal momentum of the jet going into the opposite direction 
(see Fig.~\ref{fig:ema}(b)) one is 
able to reconstruct the energy of the initial muon. Of course these cross 
section are small. Nevertheless it is worthwhile to study how these processes 
can be used and work on this is in progress.

\subsection{Bremsstrahlung from produced pairs: The ``glowing of heavy ion
collisions''}
\begin{figure}
\begin{center}
\psfig{figure=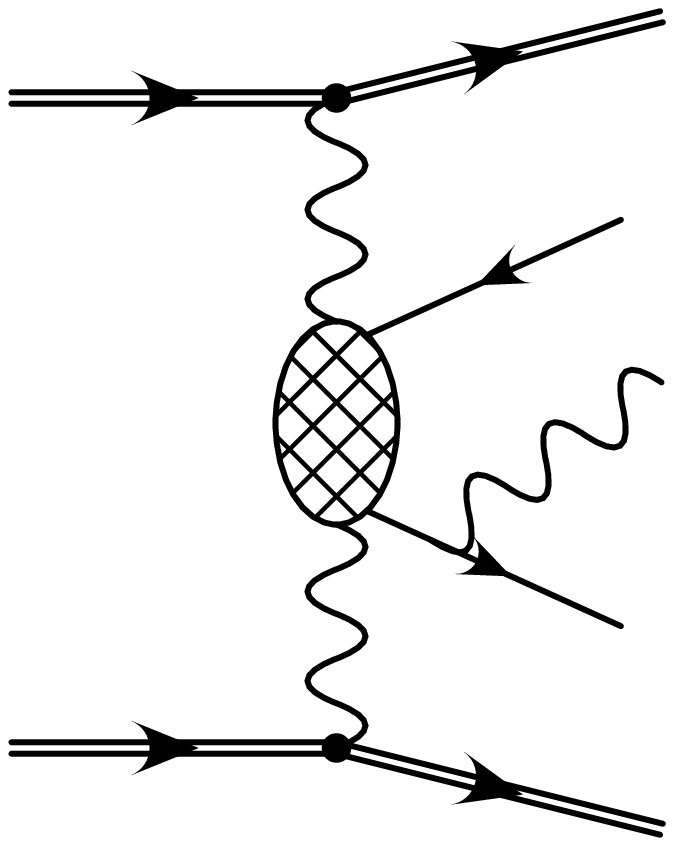,width=0.15\hsize}(a)~~~~
\psfig{figure=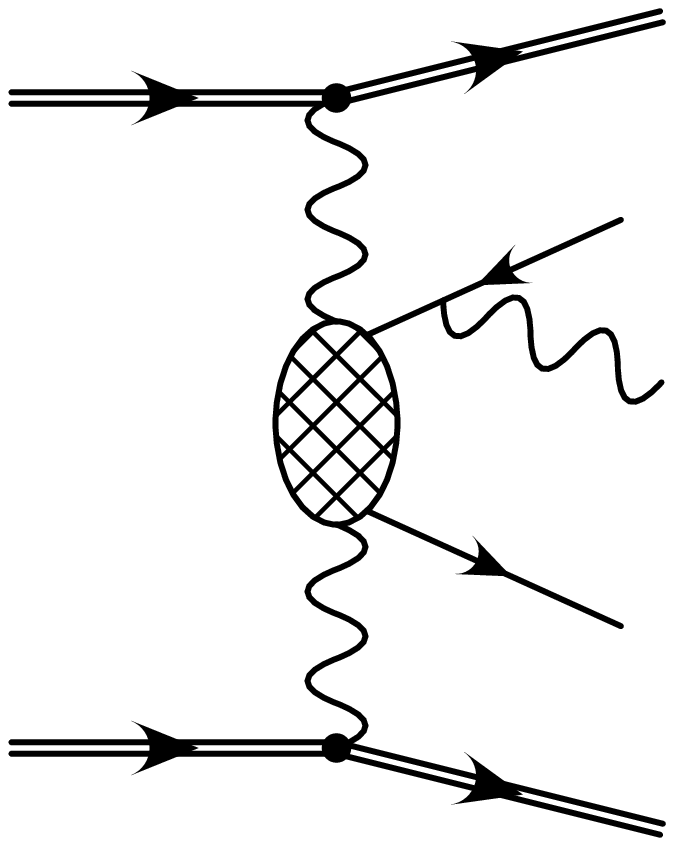,width=0.15\hsize}(b)~~~~
~~~
\psfig{figure=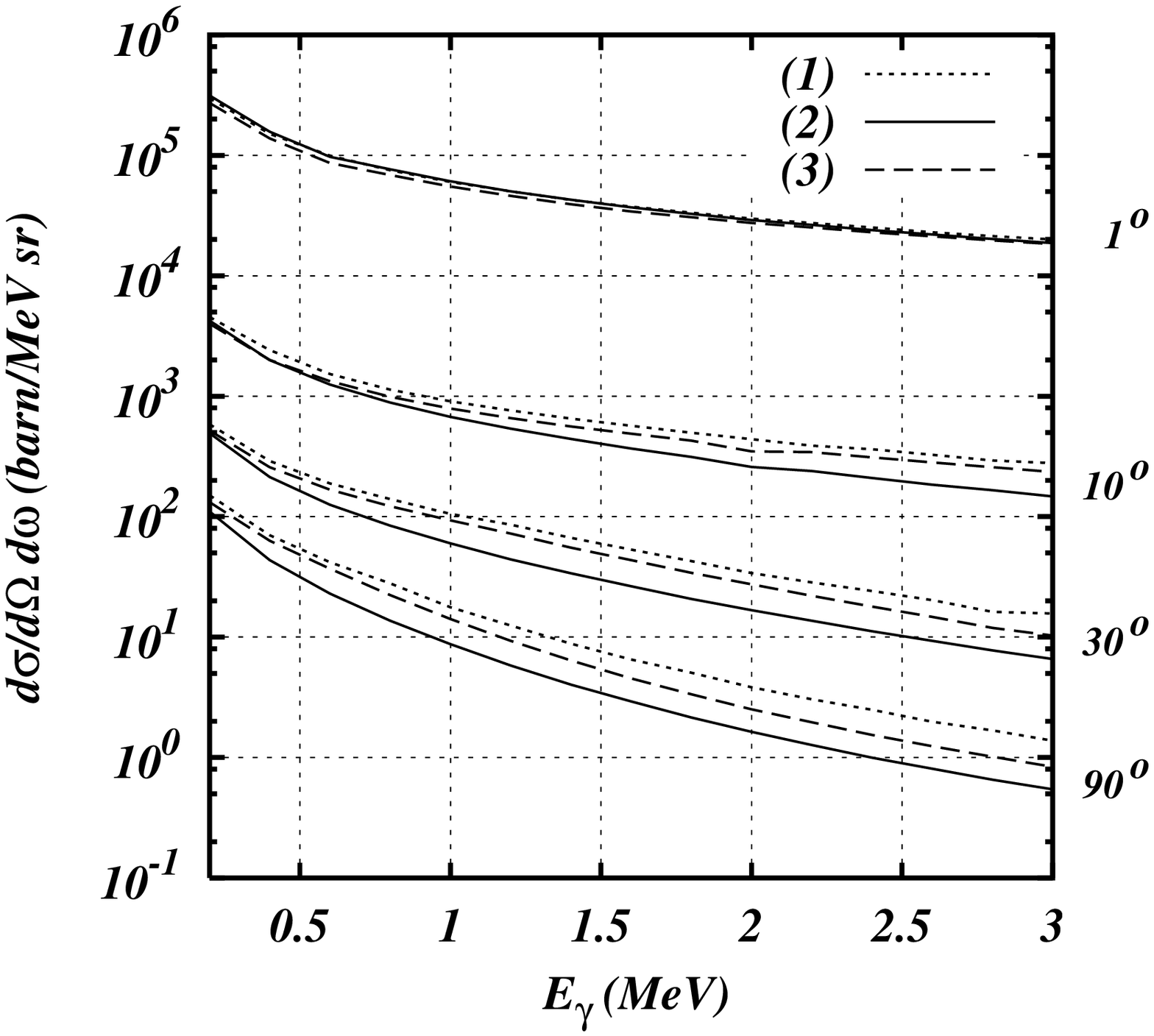,width=0.33\hsize}(c)
\end{center}
\caption{The emission of photons from the created electrons and positrons
is the main source of bremsstrahlung at ``high'' photon energies and large
angles. In the IR limit diagrams (a) and (b) give the main contribution.
The result of a calculation using three different approximation
schemes are shown for the conditions of Pb-Pb collisions at the LHC in (c).
\label{fig:brems}}
\end{figure}

The emission of bremsstrahlung photons from the ions itself is highly
suppressed due to their large mass~\cite{MeierHTB98}.
On the other hand the pair production cross section is rather large and due 
to their small mass bremsstrahlungsemission is possible easily. It is well 
known that in the case of lepton colliders this process can 
become the dominant source for photons at large angles~\cite{FadinK73}. 
A calculation of the bremsstrahlungsspectrum was done using
different approximations~\cite{HenckenTB99b}. The results for different
energies and angles are shown in Fig.~\ref{fig:brems}(c). It is an open 
question whether
these processes could be used, e.g., in order to ``see'' the interaction
vertex of the two ion bunches directly by looking at these photons.

\section{Summary}
The strong electromagnetic
fields can be used for both photon-photon and photon-ion processes. These 
processes are measurable in very peripheral collisions, as only the long 
range electromagnetic interaction is present in this case. They are 
interesting processes to be studied at both RHIC and LHC.
The theoretical description of these processes is well understood and 
corrections to the dominant coherent photon emission have been studied already.
Electron-positron pair production plays a special role in these collisions.
Besides the theoretical interest connected to the strong field effects, they
are also of practical importance as a possible loss process or even
as a possibility to measure the parton distributions inside the ions.

As can also be seen from the other contributions to this workshop, the
beautiful results of the STAR detector at RHIC have shown that we have
definitely left the regime of theoreticans dreams and hopes and are entering
the aera of experimental realities and results.

\section*{Acknowledgments}
The author would like to thank Sebastian White and Bill Marciano for the
organization of this workshop and the possibility to participate. The wonderful
atmosphere has lead to a number of most interesting discussions, which will
be surely very fruitful for the future of this kind of physics.


\section*{References}
\begin{thebibliography}{99}
\bibitem{BertulaniB88}
C.~A. Bertulani and G.~Baur, 
\Journal{\PREP}{163}{299}{1988}.
\bibitem{KraussGS97}
F.~Krauss, M.~Greiner, and G.~Soff, 
\Journal{\PPNP}{39}{503}{1997}. 
\bibitem{BaurHT98}
G. Baur, K. Hencken, and D. Trautmann, Topical Review 
\Journal{\JPG}{24}{1657}{1998}
\bibitem{BaurHTS01}
G. Baur, K. Hencken, D. Trautmann, S. Sadovsky, and {Yu}. Kharlov,
to appear in \PREP, 2002, e-print hep-ph/0112211.
\bibitem{Fermi24}
E. Fermi, \Journal{\ZP}{29}{315}{1924}.
\bibitem{Weizsaecker34}
C.~F. Weizs{\"a}cker, \Journal{\ZP}{88}{612}{1934}.
\bibitem{Williams34}
E.~J. Williams,\Journal{\PR}{45}{729}{1945}.
\bibitem{BudnevGM75}
V.~M. Budnev, I.~F. Ginzburg, G.~V. Meledin, and V.~G. Serbo,
\Journal{\PREP}{15}{181}{1975}.
\bibitem{herafuture96}
G.~Ingelman, A.~{De Roeck}, and R.~Klanner, eds.,
{\em Future Physics at HERA}, 1996.
\bibitem{HenckenTB96}
K.~Hencken, D.~Trautmann, and G.~Baur,
\Journal{\PRC}{53}{2532}{1996}.
\bibitem{OhnemusWZ94}
J.~Ohnemus, T.~F. Walsh, and P.~M. Zerwas,
\Journal{\PLB}{328}{369}{1994}.
\bibitem{DreesZ89}
M. Drees and D.~Zeppenfeld,
\Journal{\PRD}{39}{2536}{1989}.
\bibitem{Gerhard}
G. Baur, this proceedings.
\bibitem{Carlos}
C. A. Bertulani, this proceedings.
\bibitem{Klein01b} S.~R.~Klein and the STAR collaboration, LBNL-47723,
e-print nucl-ex/0104016.
\bibitem{Kristof}
K. Piotrzkowski, this proceedings.
\bibitem{Joakim}
J. Nystrand, this proceedings.
\bibitem{Pablo}
P. Yepes, this proceedings.
\bibitem{Sebastian}
S. White, this proceedings.
\bibitem{LandauL34}
L.~D. Landau and E.~M. Lifshitz,
{\em Phys. Z. Sowjet.}, {\bf 6}, 244 (1934).
\bibitem{Racah37}
G~Racah,  \Journal{\NCA}{14}{93}{1937}.
\bibitem{Baur90}
G. Baur,
\Journal{\PRA}{42}{5736}{1990}.
\bibitem{Heitler34}
W. Heitler, {\em The Quantum Theory of Radiation},
Oxford University Press, London, 1954.
\bibitem{HenckenTB95a}
K.~Hencken, D.~Trautmann, and G. Baur.
\Journal{\PRA}{51}{998}{1995}.
\bibitem{AlscherHT97}
A.~Alscher, K.~Hencken, D.~Trautmann, and G.~Baur,
\Journal{\PRA}{55}{396}{1997}.
\bibitem{HenckenTB95b}
K.~Hencken, D.~Trautmann, and G.~Baur,
\Journal{\PRA}{51}{1874}{1995}.
\bibitem{LeeMS01}
R.~N. Lee, A.~I. Milstein, and V.~G. Serbo,
e-print hep-ph/0108014, 2001.
\bibitem{Aste01}
A. Aste, G. Baur, K. Hencken, and D. Trautmann,
e-print hep-ph/0112193, 2001.
\bibitem{BetheM54}
H.~A. Bethe and L.~C. Maximon.
\Journal{\PR}{93}{768}{1954}.
\bibitem{DaviesBM54}
H.~Davies, H.~A. Bethe, and L.~C. Maximon.
\Journal{\PR}{93}{788}{1954}.\
\bibitem{IvanovSS99}
D.~{Yu.} Ivanov, A.~Schiller, and V.~G. Serbo.
\Journal{\PLB}{454}{155}{1999}.
\bibitem{SegevW98}
B.~Segev and J.~C. Wells,
\Journal{\PRA}{57}{1849}{1998}.
\bibitem{BaltzM98}
A.~J. Baltz and L.~McLerran,
\Journal{\PRC}{58}{1679}{1998}.
\bibitem{EichmannRSG99}
U.~Eichmann, J.~Reinhardt, S.~Schramm, and W.~Greiner,
\Journal{\PRA}{59}{1223}{1999}.
\bibitem{HenckenTB99}
K.~Hencken, D.~Trautmann, and G.~Baur,
\Journal{\PRC}{59}{841}{1999}.
\bibitem{SegevW98b}
B.~Segev and J.~C. Wells,
\Journal{\PRC}{59}{2753}{1999}.
\bibitem{LeeM00}
R.~N. Lee and A.~I. Milstein,
\Journal{\PRA}{61}{032103}{2000}.
\bibitem{Klein01}
S.~R. Klein,
\Journal{\NIMA}{59}{51}{2001}.
\bibitem{MeierHHT01}
H. Meier, Z. Halabuka, K. Hencken, D. Trautmann, and G. Baur,
\Journal{\PRA}{63}{032713}{2001}.
\bibitem{ChenZ75}
M.~Chen and P. Zerwas,
\Journal{\PRD}{12}{187}{1975}.
\bibitem{BaierFK73}
V.~N. Baier, V.~S. Fadin, and V.~H. Khoze,
\Journal{\NPB}{65}{381}{1973}.
\bibitem{MeierHTB98}
H. Meier, K. Hencken, D. Trautmann, and G. Baur,
\Journal{\EPJC}{2}{741}{1998}.
\bibitem{FadinK73}
V.~S. Fadin and V.~A. Khoze,
{\em Sov. Phys.-JETP} {\bf 17}, 313 (1973).
\bibitem{HenckenTB99b}
K.~Hencken, D.~Trautmann, and G.~Baur,
\Journal{\PRC}{60}{34901}{1999}.
\bibitem{BlankenbeclerD87}
 R.~Blankenbecler and S.D.~Drell, 
\Journal{\PRD}{36}{2846}{1987}.
\end{thebibliography}
\end{document}